\documentclass[12pt]{article}
\usepackage[latin1]{inputenc}
\usepackage{pstricks}
\usepackage{amsmath}
\usepackage{float}
\usepackage{color}
\input epsf
\usepackage{amssymb}
\usepackage[dvips]{graphicx,psfrag}
\newcommand{\be}{\begin{equation}}
\newcommand{\ee}{\end{equation}}
\newcommand{\bea}{\begin{eqnarray}}
\newcommand{\eea}{\end{eqnarray}}
\newcommand{\lb}{\label}
\begin{document}
\begin{titlepage}
\title{The growth of matter perturbations in some scalar-tensor DE models}
\author{Radouane Gannouji\thanks{email:gannouji@lpta.univ-montp2.fr}~
  and David Polarski\thanks{email:polarski@lpta.univ-montp2.fr}\\ 
\hfill\\
Lab. de Physique Th\'eorique et Astroparticules, CNRS\\ 
Univ. Montpellier II, France}
\pagestyle{plain}
\date{\today}

\maketitle

\begin{abstract}
We consider asymptotically stable scalar-tensor dark energy (DE) models for which 
the equation of state parameter $w_{DE}$ tends to zero in the past. The viable 
models are of the phantom type today, however this phantomness is milder than in 
General Relativity if we take into account the varying gravitational constant when 
dealing with the SNIa data. 
We study further the growth of matter perturbations and we find a scaling behaviour 
on large redshifts which could provide an important constraint. 
In particular the growth of matter perturbations on large redshifts in our 
scalar-tensor models is close to the standard behaviour $\delta_m \propto a$, 
while it is substantially different for the best-fit model in General Relativity 
for the same parametrization of the background expansion. 
As for the growth of matter perturbations on small redshifts, we show that in 
these models the parameter $\gamma'_0\equiv \gamma'(z=0)$ can take absolute values 
much larger than in models inside General Relativity. Assuming a constant $\gamma$ 
when $\gamma'_0$ is large would lead to a poor fit of the growth function $f$.
This provides another characteristic discriminative signature for these models. 
\end{abstract}

\end{titlepage}

\section{Introduction}
The aim of Dark Energy models is to explain the late-time accelerated 
expansion of the universe \cite{P97}. Like for inflationary models producing an 
early stage of accelerated expansion, we have now a wide variety of DE 
models that can account for the late-time background evolution \cite{SS00}. 
In the same way that inflationary models are constrained by the cosmological 
perturbations they produce, DE models can be constrained by the background 
evolution and their effect on the growth of perturbations. 
In principle what is basically needed is a smooth component with a sufficiently 
negative pressure. Among DE models, $\Lambda$CDM, although it contains a 
cosmological constant which can be seen as ``unnaturally'' small, is the 
simplest model and this model is presently in good agreement with observations 
on large scales (see however e.g.\cite{Perc07}). 
Another class of appealing models are quintessence models containing a minimally 
coupled scalar field \cite{RPW88}. A possible drawback of these models is their 
inability to violate the weak energy condition and to account for a phantom phase.   
Many more models of ever increasing sophistication have been proposed \cite{D07}. 
To make progress, it is important to find ways to select those classes of DE models that 
remain observationnally viable and to find characteristic signatures that will 
enable us to constrain them, or even rule them out, with more accurate data  
at our disposal in the future. 

In particular, an interesting family of DE models are those where gravity is no longer 
described by General Relativity (GR). Indeed there has been considerable interest recently 
in DE models with gravitation modified with respect to General Relativity, a feature that 
is quite generic in higher dimensional theories or also in the low energy effective action 
of more fundamental four-dimensional theories. Well-motivated models belonging to this class 
which can be explored thoroughly are scalar-tensor dark energy models 
\cite{ST,BEPS00,EP01,GPRS06,MSU06}.
Like the usual quintessence models containing a minimally coupled scalar field, scalar-tensor 
models have an additional physical degree of freedom, namely the scalar partner of the 
graviton. However, these models are more complicated as they have two free fundamental 
functions in their lagrangian, one more function in addition to the scalar field potential. 
This additional function reflects the modification of gravity encoded in the theory. 
A generic feature of these models, like for essentially all alternatives to the 
cosmological constant model, is that DE has a time-varying equation of state. 
The fact that $\Lambda$CDM fits well a large amount of observations should be an  
incentive to look for models with time-varying equation of state still able to compete 
with $\Lambda$CDM. 

As for all DE models, scalar-tensor DE models are characterized by the accelerated 
expansion they produce at low redshifts but this background effect is common to all 
DE models. The modification of gravity with respect to General Relativity is more 
specific and it expresses itself in particular in the modified growth of linear 
cosmological matter perturbations. In this way, under quite general asumptions, it 
could be possible to determine from the growth of matter perturbations combined 
with the background expansion whether a DE model lies inside General Relativity or 
not \cite{S98,B06} (see also \cite{KST06}).
Hence it is very important to investigate how this can be extracted from the 
observations \cite{HKV07,HL06}. 
Scalar-tensor DE models are certainly good examples to investigate this issue. 
Aside from deeper theoretical motivations, an additional incentive to consider 
scalar-tensor models could come from the observations if these support DE models 
which have a phantom phase (see also \cite{FWZ05} for other models that can 
produce a phantom phase). 

We will consider asymptotically stable, internally consistent, scalar-tensor DE models \cite{GPRS06}. 
For such models $F\to F_{\infty}$= constant and asymptotic stability is 
possible for $w_{DE}\to 0$ and $\Omega_m\to \Omega_{m,\infty}$. We can 
consider varying equations of state $w_{DE}(z)$ provided $w_{DE}\to 0$.
As we will show these models exhibit a characteristic signature in the growth of matter 
perturbations on large and on small redshifts. This could potentially allow us to 
constrain them or even rule them out. In connection with the growth of matter perturbations 
on small redshifts, these models will illustrate results derived in an earlier work \cite{GP07}, 
namely the possibility to have a parameter $\gamma'_0\equiv \frac{d\gamma}{dz}(z=0)$ 
whose absolute value is much larger than in General Relativity when we write the growth 
factor as $f \simeq \Omega_m^{\gamma}$. 
Indeed, it was shown in \cite{GPRS06} that $|\gamma'_0|\lesssim 0.02$ in 
models with a constant or a (smoothly) varying equation of state inside GR 
and hence also for $\Lambda$CDM. 
On the other hand, on large redshifts there can also be an important effect on the growth 
of perturbations. Interestingly, as we will 
see, we can have models for which the growth of matter perturbations on large scales is close 
to that in $\Lambda$CDM and for which the growth of matter perturbations deviates most from 
that in $\Lambda$CDM on small redshifts. 
Another interesting aspect of a varying gravitational constant is its effect on the 
interpretation of SNIa data and we will show that less phantomness is required today 
by the observations for our models compared to General Relativity. Indeed as we consider 
DE models where $w_{DE}\to 0$ in the past, a large amount of phantomness is required by 
the observations if we are inside GR while in our models this is much less the case. 
%
%

\section{An asymptotically stable scalar-tensor model}
It is important to review here the basic aspects and definitions of our 
scalar-tensor models. 
We consider DE models where gravity is described by the Lagrangian 
density in the Jordan (physical) frame 
\begin{equation}
L=\frac{1}{2} \Bigl (F(\Phi)~R -
Z(\Phi)~g^{\mu\nu}\partial_{\mu}\Phi\partial_{\nu}
\Phi \Bigr) - U(\Phi) + L_m(g_{\mu\nu})~.
\label{L}
\end{equation}
The Brans-Dicke parameter is given by $\omega_{BD}= \frac{F \Phi'^2}{F'^2}$ 
where a prime denotes a derivative with respect to redshift $z$, 
and 
\be
G_{N}= (8\pi F)^{-1}~.
\ee
Specializing to a spatially flat universe,
the DE energy density and pressure are {\it defined} as follows
\bea
3F_0~H^2  &=& \rho_m + \rho_{DE} \lb{E1a} \\
-2F_0~{\dot H} &=&  \rho_m + \rho_{DE} + p_{DE} ~. \lb{E2a}
\eea
With these definitions, the usual conservation equation applies:
\be
{\dot \rho_{DE}} = -3H ( \rho_{DE} + p_{DE} )~.
\ee
With the equation of state parameter $w_{DE}$ defined through
\be
w_{DE} \equiv \frac{p_{DE}}{\rho_{DE}}~,
\ee
the time evolution DE obeys the usual rule
\be
\frac{\rho_{DE}(z)}{\rho_{DE,0}}
\equiv \epsilon (z) = \exp \left[ 3\int_{0}^z
dz'~\frac{1+w(z')}{1+z'}\right]~.\lb{gz}
\ee
Equations (\ref{E1a}) can be rewritten as
\be 
h^2(z)= \Omega_{m,0} ~(1+z)^3 + \Omega_{DE,0} ~\epsilon(z)~,\lb{hz}
\ee
where $\Omega_{DE,0}= 1 -\Omega_{m,0}$ by definition as we assume 
a spatially flat universe. For these models
\be
\rho_{DE} + p_{DE} = \dot \Phi^2 +
{\ddot F} - H {\dot F} + 2(F - F_0)~{\dot H}~,\lb{wec}
\ee
hence the weak energy condition for DE can be violated (\cite{GPRS06}, see also \cite{T02}).

We consider viable models satisfying the following requirements:
\par\noindent
1) $F\to F_{\infty}= {\rm constant} < F_0$ for $z\to \infty$.
\par\noindent
2) The DE equation of state evolves according to $w_{DE}\to 0$ for $z\to \infty$.    
\par\noindent
3) Consistency requires $\phi'^2 \equiv \frac{3}{4}\left(\frac{F'}{F}\right)^2 + 
\frac{\Phi'^2}{2F} > 0$. 
\par\noindent
4) We impose $\omega_{BD,0}> 4 \times 10^4$.   

Condition 1) is reminiscent of the models considered in \cite{LB07}.
For these models we have that  
$\frac{G_{\rm eff}(z)}{G_{N,0}}\to \frac{G_{{\rm eff},\infty}}{G_{N,0}}$ = constant 
from some redshift on, and different from one by a few percents only. 
In order to have accelerated expansion at the present time we need some dynamical 
DE equation of state.   
We consider a parametrization \cite{CP01},\cite{L03} with a smoothly varying equation 
of state where DE tends to a scaling behaviour in the past
\be
w_{DE}(z)  =  (-1 + \alpha) + \beta ~(1-x) \equiv w_0 + w_1~\frac{z}{1+z}~,\\ \lb{wCPL}
\ee
where $x\equiv \frac{a}{a_0}$. 
If we impose $w_{DE}\to 0$ for $z\to \infty$, this prameterization reduces to the 
simple form 
\be
w_{DE}(z)  = -\frac{\beta}{1+z}~.\lb{wbeta}
\ee
The corresponding DE evolution reads \cite{CP01}
\be
\epsilon (z) = (1+z)^{3} {\rm e}^{-3\beta \frac{z}{1+z}}~.\\ \lb{epsCPL}
\ee
For given cosmological parameters $\Omega_{DE,0}$ and $\Omega_{m,0}$
the background evolution is completely fixed by the parameter $\beta$.
We have in particular 
\be
\Omega_{m}(z) = \left[e^{-3\beta\frac{z}{1+z}} \frac{\Omega_{DE,0}}{\Omega_{m,0}} 
                                              + 1\right]^{-1}~.\lb{Omz}
\ee
Clearly, $\Omega_m\to \Omega_{m,\infty}(\beta)$ 
\be 
\Omega_{m,\infty}(\beta) = \left[e^{-3\beta} \frac{\Omega_{DE,0}}{\Omega_{m,0}} 
                                              + 1\right]^{-1}~.\lb{Ominf}
\ee
The quantity $\Omega_{m}(z)$ is fixed by $\Omega_{m,0}$ and $\beta$. 
As shown in \cite{GPRS06}, the requirement 1) implies the inequality (independent 
of the specific form of $F(z)$)
\be
\frac{F_{\infty}}{F_0} > \Omega_{m,\infty}~.\lb{condF}
\ee
In particular a large amount of phantomness today implies $\Omega_{m,\infty}$ 
close to one 
and hence $F_{\infty}$ close to $F_0$.
We use further the following ansatz
\be
\frac{F}{F_0}(z) =\frac{F_{\infty}}{F_0}+(1-\frac{F_{\infty}}{F_0})
                        \left(\frac{5}{(1+z)^4}-\frac{4}{(1+z)^{5}}\right)~.\lb{ansF}
\ee 
This ansatz satisfies exactly $F_1 \equiv \frac{F'}{F_0}(z=0) =0$ while $F_2 \equiv \frac{F''}{2 F_0}(z=0) 
=-10 (1-\frac{F_{\infty}}{F_0})<0$.
For the ansatz (\ref{ansF}), we must have in addition $\frac{F_{\infty}}{F_0} < 1 + 
\frac{3}{20} \Omega_{DE,0} (1-\beta)$ which comes from the requirement 
$\phi'^2>0$ at $z=0$. So we impose the functions $h(z)$ and $\frac{F}{F_0}(z)$ 
from which all other quantities can be reconstructed and we check the physical 
consistency for each reconstructed scalar-tensor model. As $h(z)$ is imposed, so 
is the background dynamics. In this way we can compare different DE models inside 
and outside General relativity possessing the same background evolution.
Clearly, as the DE equation of state parameter $w_{DE}$ tends to 
zero as $z$ increases, it must start being phantom today if it is to pass the 
observational constraints. This can be seen more quantitatively using the 
constraint on the shift parameter. We see from Figure \ref{R} that the viable models 
are of the phantom type today ($\beta>1$). The possibility to have a phantom DE 
sector today is actually an attractive feature of scalar-tensor DE models and is 
not excluded by the observations.
\section{Some observational constraints}
 
To get some insight into the parameter window for viable models, we constrain them 
using Supernovae data, BAO (Baryonic acoustic oscillations) data and CMB data. 
We have to maximize the probability function 
\be
P(\Omega_{m,0},\beta) \propto e^{-\frac{1}{2}~\chi^2}~.
\ee
where $\chi^2=\chi^2_{\rm{SN}} + \chi^2_{A} + \chi^2_{R}$ for a background evolving according 
to eqs.(\ref{hz}),(\ref{epsCPL}). We use a sample consisting of 192 Supernovae \cite{D07,R07} 
for which
\be
\chi^2_{\rm SN} = \sum_{i=1}^{192}\frac{(\mu_{th,i}-\mu_{exp,i})^2}{\sigma_i^2}~,
\ee
with 
\be
\mu_{th,i} = 5 \log\left((1+z_i)\int_0^{z_i}\frac{{\rm d}z}{h}\right)+ 
               \mu_0+\frac{15}{4}~\log~\frac{G_{\rm{eff}}(z_i)}{ G_{\rm{eff},0}}~,\lb{mu}
\ee
where $\mu_0 = 25 + 5~\log~\left(\frac{cH_0^{-1}}{{\rm Mpc}}\right)$, the distance 
modulus $\mu$ is the difference between the apparent magnitude $m$ and the absolute 
magnitude $M$. The important quantity $G_{\rm{eff}}$ is defined as 
\be
G_{\rm eff} = G_N ~\frac{F+2(dF/d\Phi)^2}{F+\frac{3}{2}(dF/d\Phi)^2}
                = G_N ~\frac{1 + 2 \omega^{-1}_{BD}}{1 + \frac{3}{2} \omega^{-1}_{BD}} ~.\lb{GeffST}
\ee
We have for $\omega_{BD}\gg 1$  
\be
G_{\rm eff} \simeq  G_N ~( 1 + \frac{1}{2} \omega^{-1}_{BD})~.
\ee
In particular $G_{\rm{eff},0} \simeq  G_{N,0}$ due to the well-known strong 
solar system gravitational constraint $\omega_{BD,0}>4\times 10^4$.
We get rid of the nuisance parameter $H_0$ using the simple way suggested 
by \cite{PC03}, integrating over $H_0$ gives essentially the same result. 
Note the addition of the last term in eq.(\ref{mu}) which takes into account a varying 
gravitational constant \cite{RU02}. This term allows to discriminate different scalar-tensor models 
using SNIa data even for similar background expansion. We will come back to this point 
below.    

The BAO constraints can be expressed as a constraint on the quantity $A$
\be
A(z) = \frac{\sqrt{\Omega_{m,0}}}{z}~\left[\frac{z}{h(z)}~
                  \left(\int_0^z {\rm d} z' \frac{1}{h(z')}\right)^2\right]^{\frac{1}{3}}~.
\ee
with \cite{E05}
\be
A = 0.469 \pm 0.017 ~,
\ee
and
\be
\chi^2_A = \frac{\left( A(z=0.35,\Omega_{m,0},\beta)-0.469 \right)^2}{(0.017)^2}~.
\ee
We have finally a constraint on the shift parameter extracted from the CMB data
\be
R = \sqrt{\Omega_{m,0}} \int_0^{1089} \frac{{\rm d}z}{h(z)}~,\lb{shift}
\ee
with \cite{WM06}
\be
R = 1.70 \pm 0.03 ~,
\ee
and
\be
\chi^2_R = \frac{\left(R(\Omega_{m,0},\beta) - 1.7\right)^2}{(0.03)^2}~. 
\ee
As we can see from Figure 1, the shift parameter constrains our model to be 
of the phantom type today, $\beta>1$. This is expected because in our model 
the equation of state of DE tends to that of dust in the past. If we remember 
that a cosmological constant agrees fairly with the data, our model must 
compensate by being phantom on small redshifts.  
\begin{table}

\begin{tabular}{ c c c }
\begin{tabular}{| c | c | c |}
\hline
$F_{\infty}/F_0$&$\Omega_{m,0}$&$\beta$\\
\hline
\hline
0.93 & $0.31^{+0.04}_{-0.04}$ & $1.09^{+0.12}_{-0.13}$ \\ 
$ $ & $ $ & $ $ \\
0.94 & $0.30^{+0.04}_{-0.03}$ & $1.12^{+0.12}_{-0.13}$ \\ 
$ $ & $ $ & $ $ \\
0.95 & $0.30^{+0.04}_{-0.03}$ & $1.15^{+0.13}_{-0.13}$ \\
$ $ & $ $ & $ $ \\
0.96 & $0.29^{+0.04}_{-0.03}$ & $1.18^{+0.12}_{-0.14}$ \\
$ $ & $ $ & $ $ \\
 GR & $ 0.27^{+0.04}_{-0.03}$ & $1.28^{+0.17}_{-0.15}$ \\
\hline
\hline
\end{tabular} &

\begin{tabular}{| c | c | c |}
\hline
$\Omega_{m,\infty}$ & $C$ & $p_1$ \\
\hline
\hline
$0.92^{+0.04}_{-0.04}$ & $0.99^{+0.04}_{-0.04}$ & $0.99^{+0.03}_{-0.03}$ \\
$ $ & $ $ & $ $\\
$0.93^{+0.04}_{-0.04}$ & $0.98^{+0.04}_{-0.04}$ & $0.99^{+0.02}_{-0.02}$ \\
$ $ & $ $ & $ $\\
$0.93^{+0.04}_{-0.04}$ & $0.98^{+0.04}_{-0.04}$ & $0.99^{+0.02}_{-0.02}$ \\
$ $ & $ $ & $ $\\
$0.93^{+0.03}_{-0.04}$ & $0.97^{+0.04}_{-0.04}$ & $0.98^{+0.02}_{-0.02}$ \\
$ $ & $ $ & $ $\\
$0.95^{+0.04}_{-0.03}$ & $0.95^{+0.04}_{-0.03}$ & $0.97^{+0.02}_{-0.02}$ \\
\hline
\hline
\end{tabular} &

\begin{tabular}{| c | c |}
\hline
$\gamma_0$ & $\gamma'_0$ \\
\hline
\hline
$0.54^{+0.01}_{-0.01}$ & $-0.07^{+0.03}_{-0.02}$ \\
$ $ & $ $\\
$0.54^{+0.01}_{-0.01}$ & $-0.06^{+0.03}_{-0.02}$ \\ 
$ $ & $ $\\
$0.54^{+0.01}_{-0.01}$ & $-0.04^{+0.02}_{-0.01}$ \\ 
$ $ & $ $\\
$0.55^{+0.02}_{-0.01}$ & $-0.03^{+0.05}_{-0.01}$ \\ 
$ $ & $ $\\
$0.56^{+0.01}_{-0.01}$ & $0.01^{+0.002}_{-0.002}$ \\
\hline
\hline
\end{tabular}\\
\end{tabular}
\caption{We summarize in this table the best-fit models for given parameter 
$F_\infty$ when all data are taken into account, with $2\sigma$ errors. 
The last line corresponds to General Relativity (a constant gravitational constant). 
It is seen that a varying gravitational constant $G_{\rm eff}$ can have a 
nonnegligible effect. In particular, though the quantity $\Omega_{m,\infty}$ 
is higher in GR than in the scalar-tensor models, due to the value of 
$F_{\infty}$, $C$ is closer to $1$ in these models. Hence the growth of 
matter perturbations on large redshifts for these models, $\delta_m\propto a^{p_1}$, 
is closer to the standard one as in $\Lambda$CDM, see eqs.(\ref{C}),(\ref{p}),(\ref{p1}).}
\end{table} 

We would like now to look more closely at the effect of a varying (effective) 
gravitational constant on the measurement of luminosity distances. Let us write 
eq.(\ref{mu}) in the following way
\be 
\mu_{th,i} = [ 1 + {\cal G}(z_i)]~5 \log \left((1+z_i)\int_0^{z_i}\frac{{\rm d}z}{h}\right) 
                                   + \mu_0~,\lb{muG}
\ee
where we have introduced the quantity 
\be
{\cal G}(z) = \frac{3}{4}~\frac{ \log~\frac{G_{\rm{eff}}(z)}{G_{\rm{eff},0}} }
                         { \log \left[(1+z)\int_0^{z}\frac{{\rm d}z'}{h}\right] }~.
\ee 
We can write in full generality  
\be
\frac{G_{\rm eff}(z)}{ G_{\rm{eff},0} } \equiv 1 + \Delta(z)~.
\ee
In the models studied here, $\Delta(z)$ is positive definite and at most 
of the order of a few percents, $\Delta\lesssim 0.07$, while $\Delta(z=0)$ 
vanishes by definition. For any viable scalar-tensor model we have on very 
small redshifts 
\be
\Delta(z\approx 0) \simeq \frac{1}{2}~( \omega^{-1}_{BD} - \omega^{-1}_{BD,0} )
                      +\left( \frac{F_0}{F}-1\right) \approx 0~.\lb{DelST} 
\ee
We can give a more formal expression for $\Delta(z)$ but for our purposes it 
is not needed here. 
Whenever the quantity $\Delta$ is small, we obtain 
\be
\frac{15}{4}~\log~\frac{G_{\rm{eff}}}{ G_{\rm{eff},0}}\simeq 
                   1.63~\Delta~~~~~~~~~~~~~\Delta\ll 1~.\lb{logGeff}                  
\ee
and hence also  
\be
|{\cal G}(z\approx 0)| \simeq 0.33~\frac{\Delta}{|\log z|}\ll 1~.\lb{calG2}  
\ee 
Hence it is seen from (\ref{muG}), (\ref{DelST}), (\ref{calG2}) that the 
effect of a varying $G_{\rm eff}$ for scalar-tensor dark energy models is 
negligible on very small redshifts. We have indeed checked it with our models 
using SNIa data on redshifts $z\le 0.05$. Furthermore, it is also clear from 
(\ref{logGeff}) that this effect cannot be very large whenever $\Delta\ll 1$.

Nevertheless, as we can see from Table 1, a varying gravitational constant 
$G_{\rm eff}$, characterized in our models by the parameter $F_{\infty}$, 
can have a nonnegligible effect. In particular it is seen that our models 
with $F_{\infty}<F_0$ can fit the data with dark energy which is less of 
the phantom type today than it would have to be in the corresponding DE model 
inside GR with a background expansion of the type (\ref{hz}), (\ref{epsCPL}). 
\begin{figure}
\begin{centering}\includegraphics[scale=.3]{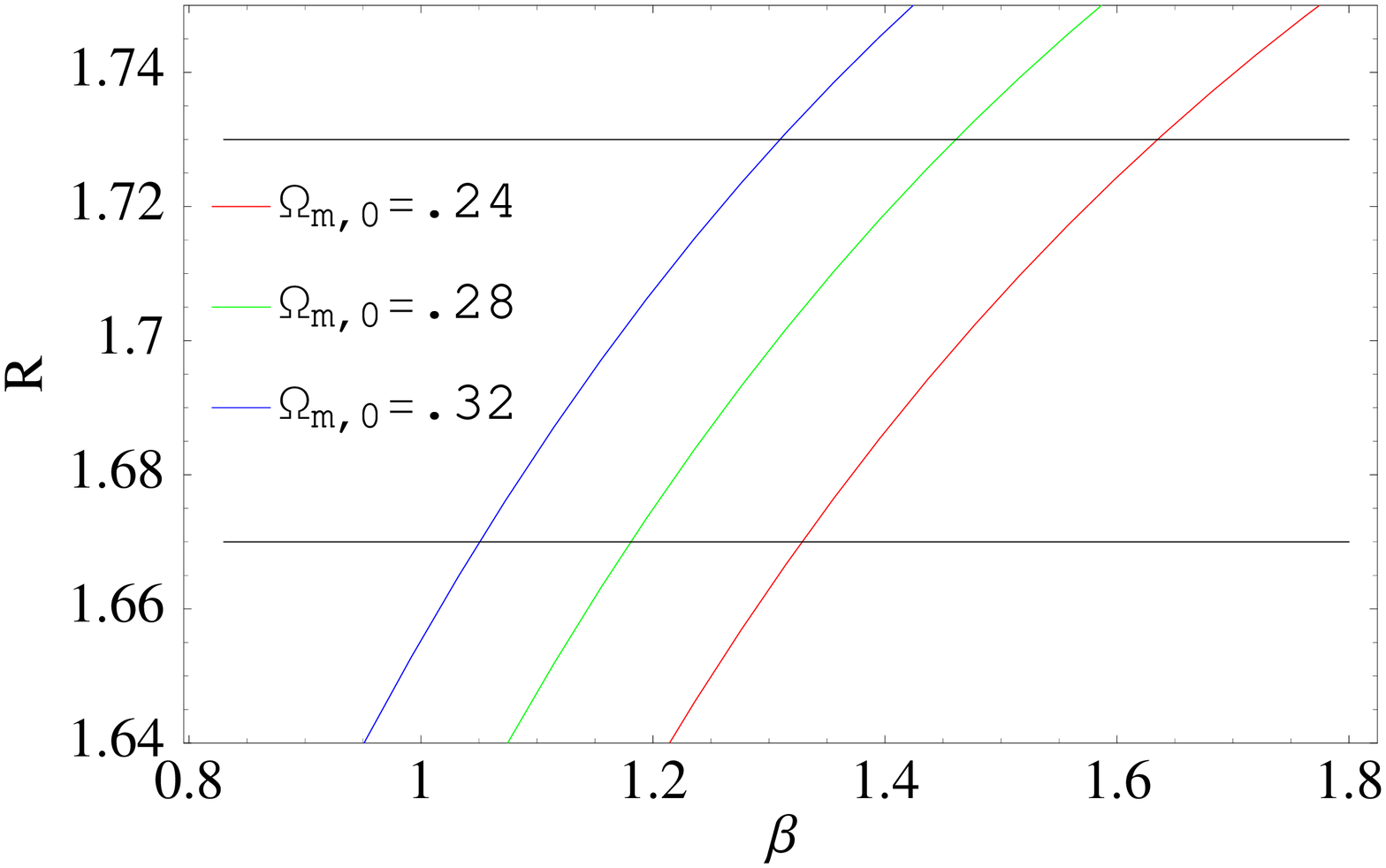}~~~\includegraphics[scale=.3]{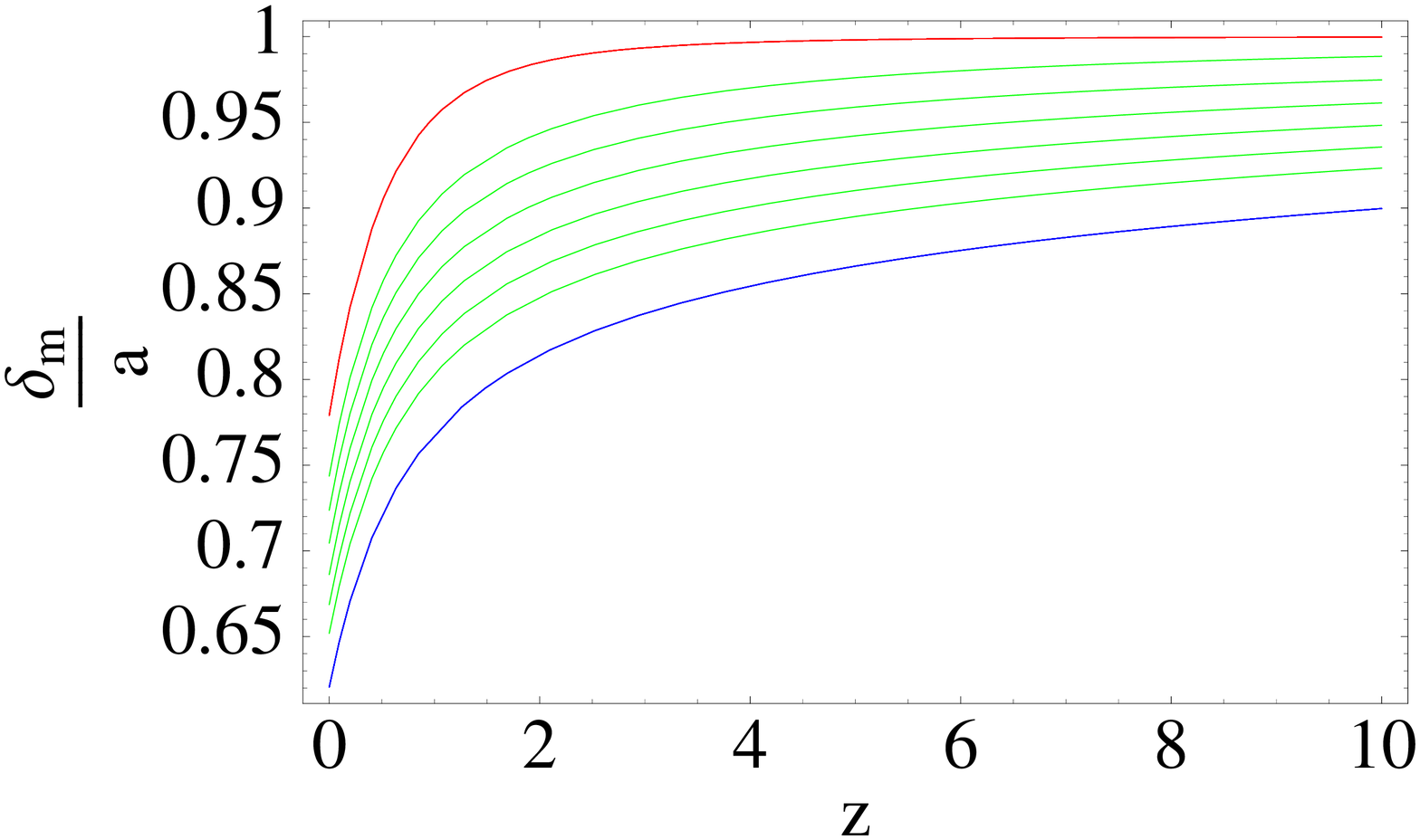}
\par\end{centering}
\caption{a) On the left, the shift parameter $R$ (see eq.(\ref{shift})) is shown 
in function of $\beta$ for several values of $\Omega_{m,0}$. The interval of 
viable models all correspond to phantom DE, $\beta>1$ which was expected because 
$w_{DE}\to 0$ in the past. 
b) On the right, the quantity $\frac{\delta_m}{a}$ (with arbitrary normalization, i.e. 
$\frac{\delta_m}{a}|_{z=100}=1$) is shown for various cases: General Relativity (GR) (blue), 
$\Lambda$CDM (red) and scalar-tensor models (green) characterized by the value of $F_{\infty}$. 
We have from top to bottom $\frac{F_{\infty}}{F_0}=0.9301, 0.94, 0.95, 0.96, 0.97, 0.98$. 
The first value corresponds to $C\approx 1$ and the linear growth on large redshifts is essentially 
similar to $\Lambda$CDM. Except for $\Lambda$CDM, all the models displayed here 
have the same background behaviour with $\Omega_{m,0}=0.3$, and $\Omega_{m,\infty}=0.93$.}
\lb{R}
\end{figure}
%
\section{Linear growth of perturbations}
Let us turn now to the dynamics of the linear matter perturbations. 
As shown in \cite{BEPS00}, these perturbations satisfy a modified equation 
of the type 
\begin{equation}
{\ddot \delta} + 2H {\dot \delta} - 4\pi G_{\rm eff}\,
\rho_m~\delta = 0~,\label{del}
\end{equation}
with $G_{\rm eff}$ given by eq.(\ref{GeffST}). 
Equation (\ref{del}) can be seen as 
a minimal modification to the growth of linear perturbations which comes from the 
modification of Poisson's equation 
\be
\frac{k^2}{a^2}~\phi = -4\pi~G~\rho~\delta  \to  
               \frac{k^2}{a^2}~\phi = -4\pi~G_{\rm eff}~\rho~\delta~.\lb{Poisson}
\ee 
This modification reflects the fact that the effective coupling constant describing 
the gravitational interaction of two close test masses is given by $G_{\rm eff}$. 
This is so on all cosmic scales of interest where the dilaton field is essentially 
massless. 
It should be stressed that the modification in (\ref{del}) is scale-independent and 
can appear in many modified gravity models (see e.g. \cite{N07}), it can even appear 
in DE models inside General Relativity if one is willing to consider DE with unusual 
properties \cite{KS07}.

It is convenient to introduce the quantity $f=\frac{d \ln \delta}{d \ln a}$, the growth 
factor of the perturbations. In function of $f$, the linear perturbations obey the 
equation 
\be
\frac{df}{dx} + f^2 + \frac{1}{2} \left(1 - \frac{d \ln \Omega_m}{dx} \right) f = 
                              \frac{3}{2} \frac{G_{\rm eff}}{G_{N,0}} \Omega_m~.\lb{fx}
\ee
with $x\equiv \ln a$. The quantity $\delta$ is recovered from $f$, 
$\delta(a) = \delta_i~{\rm exp} \left[ \int_{x_i}^{x} f(x') dx' \right]$.
We see that $f=p$ when $\delta\propto a^p$, in particular $f\to 1$ in $\Lambda$CDM for large $z$ 
while $f=1$ in an Einstein-de Sitter universe. 
In our model, $G_{\rm eff}\to G_{\rm eff,\infty}$, $\Omega_m\to \Omega_{m,\infty}$, 
these quantities tend rather quickly to their asymptotic value for $z\gg 1$. 

Introducing the quantity $C$ with  
\be
0 < C\equiv \frac{G_{{\rm eff},\infty}}{G_{N,0}} \Omega_{m,\infty} = 
                    \frac{F_0}{F_{\infty}}\Omega_{m,\infty} < 1~,\lb{C}
\ee
we see that in the asymptotic regime $G_{\rm eff}\to G_{\rm eff,\infty}=\frac{F_0}{F_{\infty}}$, 
the perturbations obey a scaling behaviour 
\be
\delta = D_1 a^{p_1} + D_2 a^{p_2}~,\lb{p} 
\ee
with
\bea
p_1 &=& \frac{1}{4} ( -1 + \sqrt{1+24 C})\lb{p1}\\
p_2 &=& \frac{1}{4} ( -1 - \sqrt{1+24 C})~.\lb{p2}
\eea
As $p_1<1$, we see that the growing mode of the perburbations allways grows \emph{slowlier} 
than in a $\Lambda$CDM universe (or in an Einstein-de Sitter universe) for $z\gg 1$. 
Therefore the amplitude of the linear matter perturbations on small redshifts before 
formation of structure starts, compared to the amplitude of perturbations derived from 
the CMB data, can be significantly different, and smaller, from that in $\Lambda$CDM. 
In particular, the perturbations will grow nonlinear on \emph{lower} redshifts, structure 
formation starts later. Further, the bias $b$ derived from $\sigma_8$ should be larger 
in these models than it is in $\Lambda$CDM.   
On the other hand, for a model for which $C$ is very close to 1 (but still satisfying 
$C<1$), both linear perturbations modes will evolve essentially like in a $\Lambda$CDM universe 
untill low redshifts where a significant departure can appear.      
Another important issue is to characterize this departure on small redshifts. 

It is well known that for a $\Lambda$CDM universe it is possible to write 
$f \simeq \Omega_m^{\gamma}$ where $\gamma$ is assumed to be constant, an 
approach pioneered in the literature some time ago \cite{P84,LLPR91}. 
There has been renewed interest lately in this approach as the growth of matter 
perturbations could be a decisive way to discriminate between models that are 
either inside or outside General Relativity (GR).
Clearly it is possible 
to write allways
\be
f = \Omega_m^{\gamma(z)}\lb{fg}~.
\ee
For $\Lambda$CDM we have $\gamma_0\equiv \gamma(z=0)\approx 0.55$.
As was shown in \cite{GP07}, for $\Lambda$CDM we have 
$\gamma'_0\equiv \frac{d\gamma}{dz}|_0\approx -0.015$. For $\Omega_{m,0}=0.3$, 
$\gamma_0=0.555$, slightly higher than the constant $\frac{6}{11}=0.5454$ derived in 
\cite{WS98} for a slowly varying DE equation of state and $\Omega_{m}\approx 1$. 
There is a very slight difference on small resdshifts $z\lesssim 0.5$ between the 
true function $f_{\Lambda}(z)$ and $\Omega_m^{\frac{6}{11}}$, one could as well 
use $\gamma=0.56$ and the agreement would be even better. As $\Omega_m\to 1$ the 
differences are important only on small redshifts. Note also that we find  
a slightly negative slope $\gamma'_0$ so that $\gamma$ comes closer to 
$\frac{6}{11}$ as $z$ increases. 
A definite departure from these values could signal a departure from a 
$\Lambda$CDM universe. More importantly as we will see later a large value for 
$\gamma'_0$ could be a hint for a DE model outside GR. This aspect was already 
emphasized in \cite{GP07}. 
We illustrate in Figure 2 the behaviour of $\gamma_0$ and $\gamma'_0$ 
in function of $\beta$ and $\Omega_{m,0}$ \emph{inside} GR. 

As emphasized in \cite{GP07}, equation (\ref{fx}) yields the following identity 
\be
\gamma'_0 = \left[ \ln \Omega^{-1}_{m,0} \right]^{-1} ~\left[ -\Omega^{\gamma_0}_{m,0} - 
   3(\gamma_0 - \frac{1}{2})~w_{{\rm eff},0} + \frac{3}{2}~\Omega^{1-\gamma_0}_{m,0} - \frac{1}{2}\right]~,
\lb{dgamma0b}
\ee 
whenever $\frac{G_{{\rm eff},0}}{G_{N,0}}=1$ to very high accuracy, which is certainly 
the case in scalar-tensor models as we have seen in Section 2. In other words we have 
a constraint 
\be
f(\gamma_0,~\gamma'_0,~\Omega_{m,0},~w_{DE,0})=0~.\lb{f0}
\ee 
As was shown in \cite{GP07}, this constraint takes the following form 
\be
\gamma'_0 \simeq -0.19 + d~(\gamma_0 - 0.5)~~~~~~~~~~~~~~~~~~~~~~~
                                                      ~d\approx 3~. \lb{dgamma0f}
\ee 
The coefficient $d$ depends on the background parameters $d=d(w_{DE,0},\Omega_{m,0})$.
For given background parameters $\Omega_{m,0}$ and $w_{DE,0}$, $\gamma'_0$ will 
take the corresponding value $\gamma'_0(\gamma_0)$. The value of $\gamma_0$ realized 
will depend on the particular model under consideration and can be obtained numerically. 
Typically we will have $\gamma'_0\ne 0$. For models inside General Relativity 
$|\gamma'_0|\lesssim 0.02$ was obtained. For example for \emph{constant} $w_{DE}$, $\gamma'_0$ 
is allmost independent of $w_{DE}=w_{DE,0}$, with $\gamma'_0\approx -0.02$ for 
$\Omega_{m,0}=0.3$, while at the same time $\gamma_0$ can have a nonnegligible variation,  
we have $0.545\lesssim \gamma_0\lesssim 0.565$ for $-1.4\lesssim w_{DE,0}\lesssim -0.8$ 
and $\Omega_{m,0}=0.3$. 
\begin{figure}
\begin{centering}\includegraphics[scale=.3]{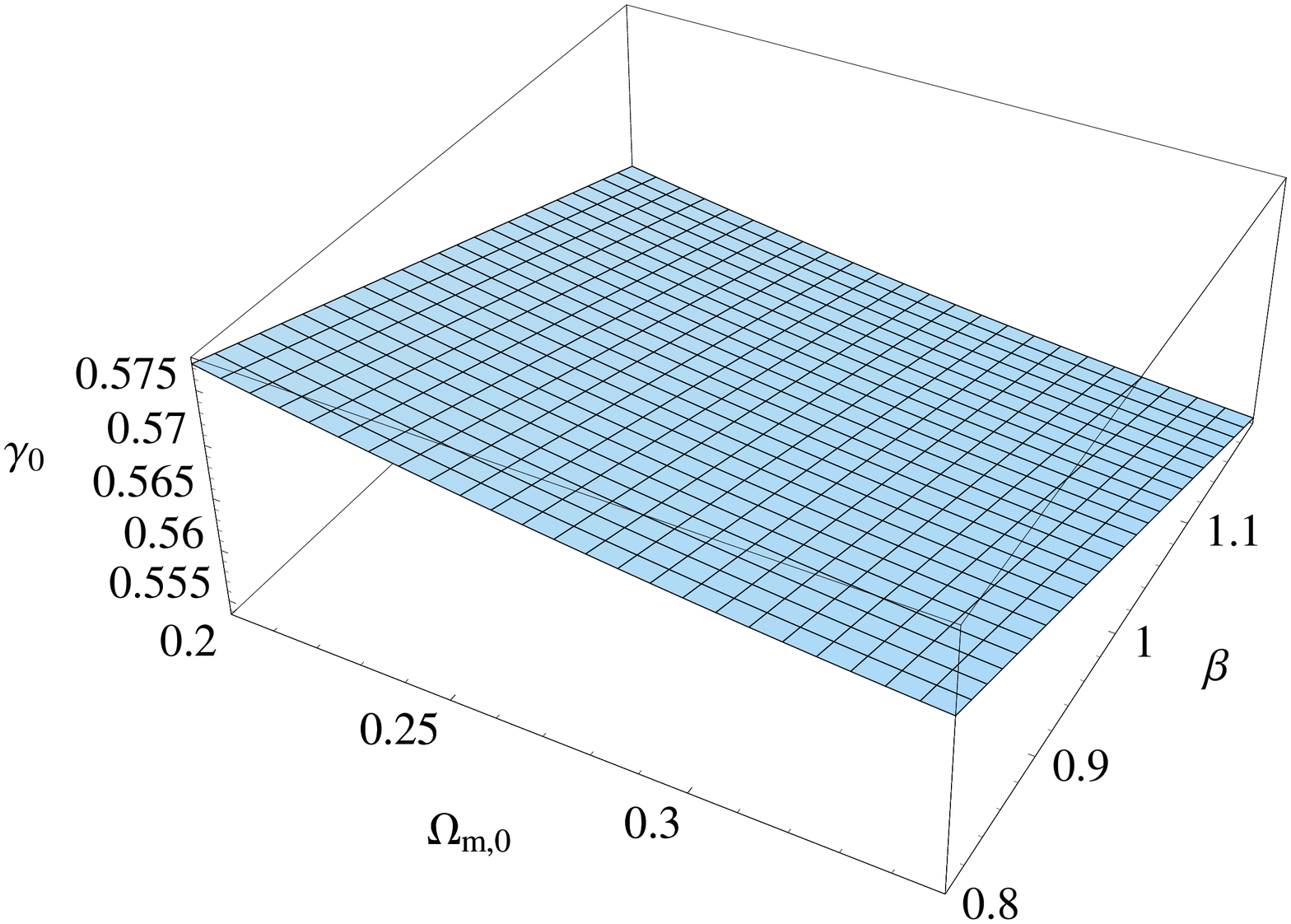}~~~\includegraphics[scale=.3]{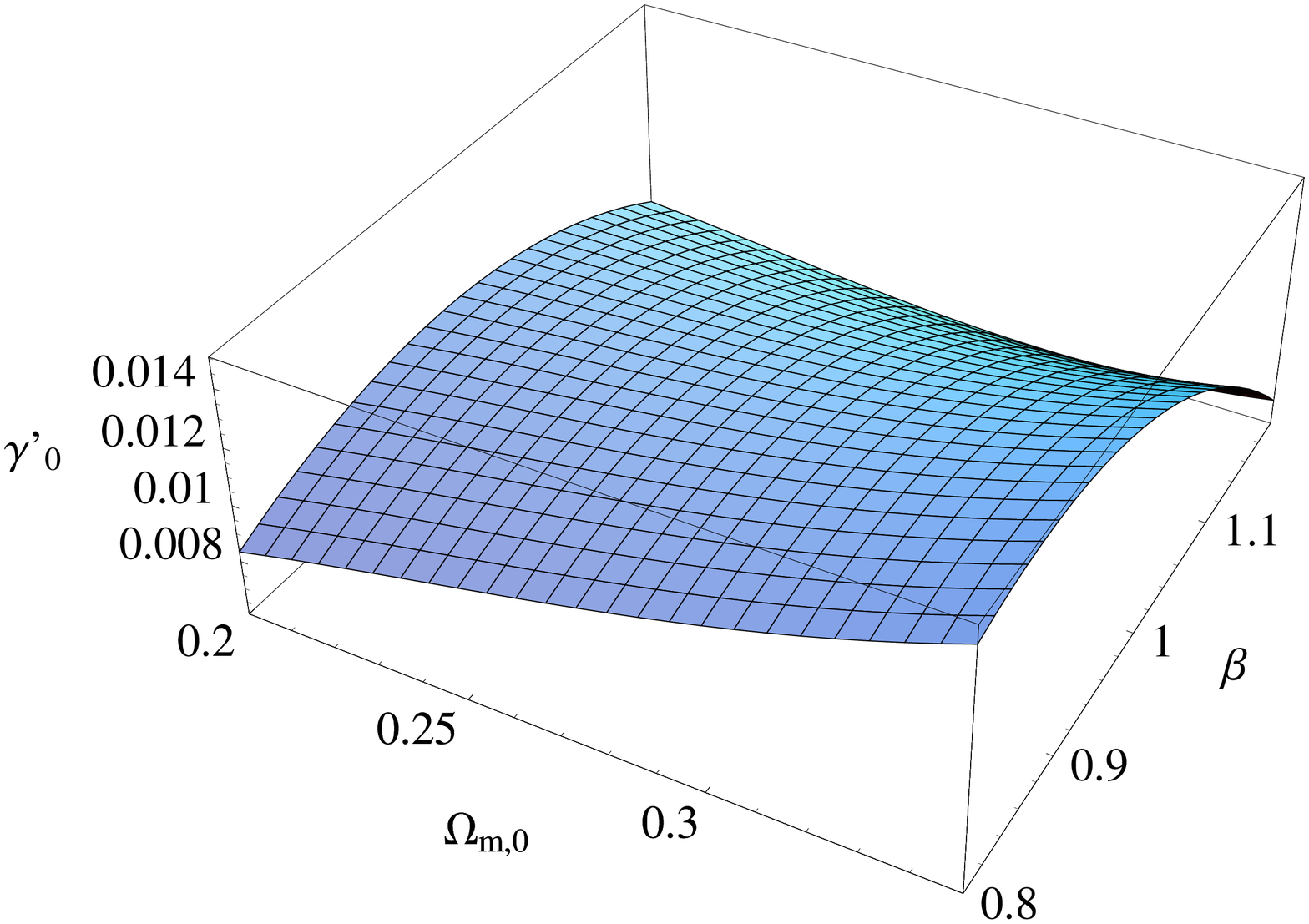} 
\par\end{centering}
\caption{a) On the left, $\gamma_0$ is shown in function of the parameters 
$\Omega_{m,0}$ and $\beta$ for the ansatz (\ref{wbeta}) \emph{assuming} General Relativity. 
b) On the right, the corresponding $\gamma'_0$ is displayed and it is seen that 
$|\gamma'_0|\lesssim 0.015$.}
\end{figure}
We can compare scalar-tensor models with different values of $F_{\infty}$ but identical 
background evolution parametrized using the parameters $\beta$ and $\Omega_{m,0}$, 
according to eqs.(\ref{hz}), (\ref{wbeta}), (\ref{epsCPL}). 
These models can be distinguished in all observations affected by a varying 
gravitational constant but they will be undistinguishable with respect to 
purely background constraints. 
In particular, they yield different perturbations growth factor $f$, or equivalently 
different $\gamma$. 
\begin{figure}
\begin{centering}\includegraphics[scale=.3]{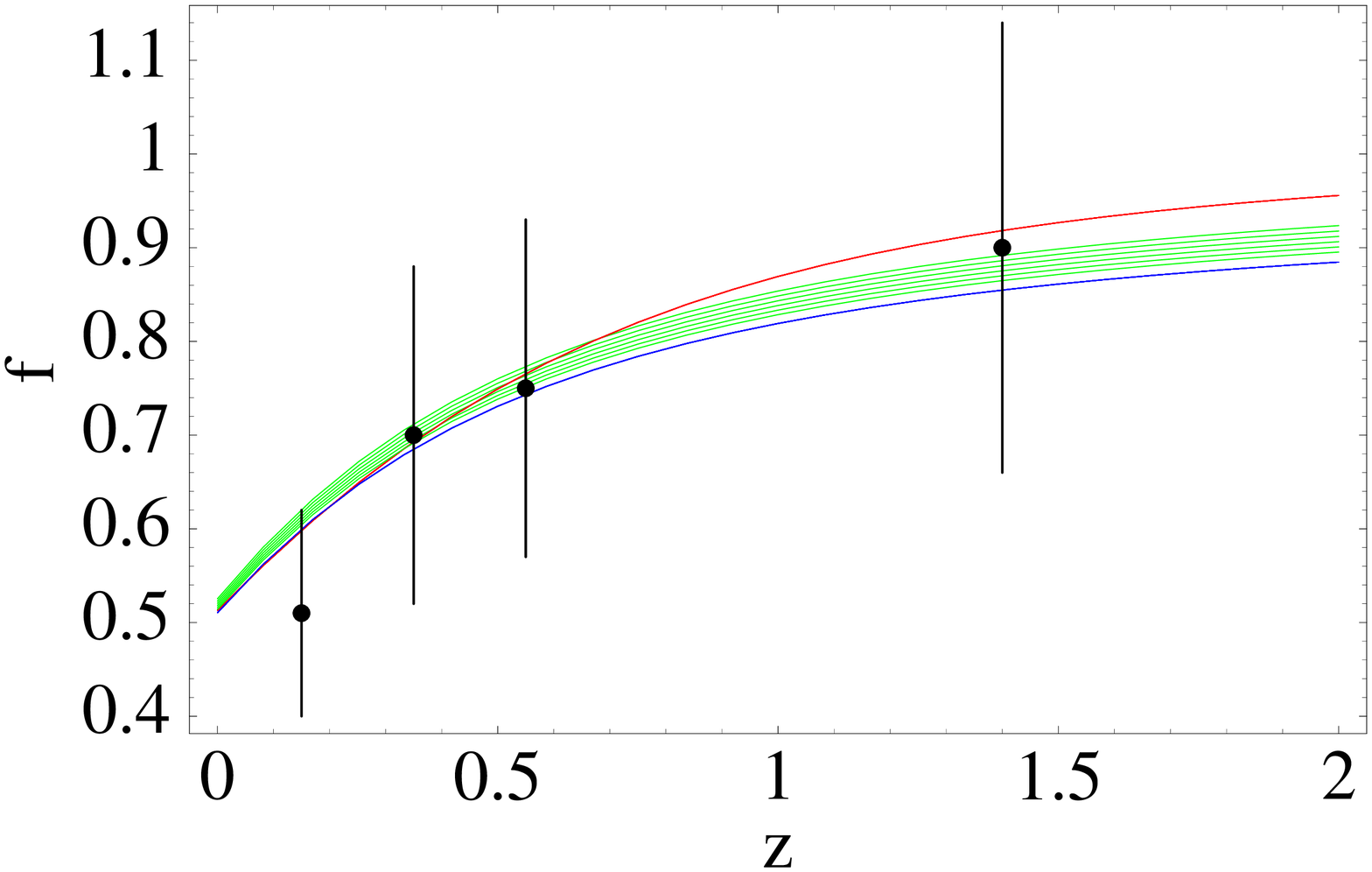}~~~\includegraphics[scale=.3]{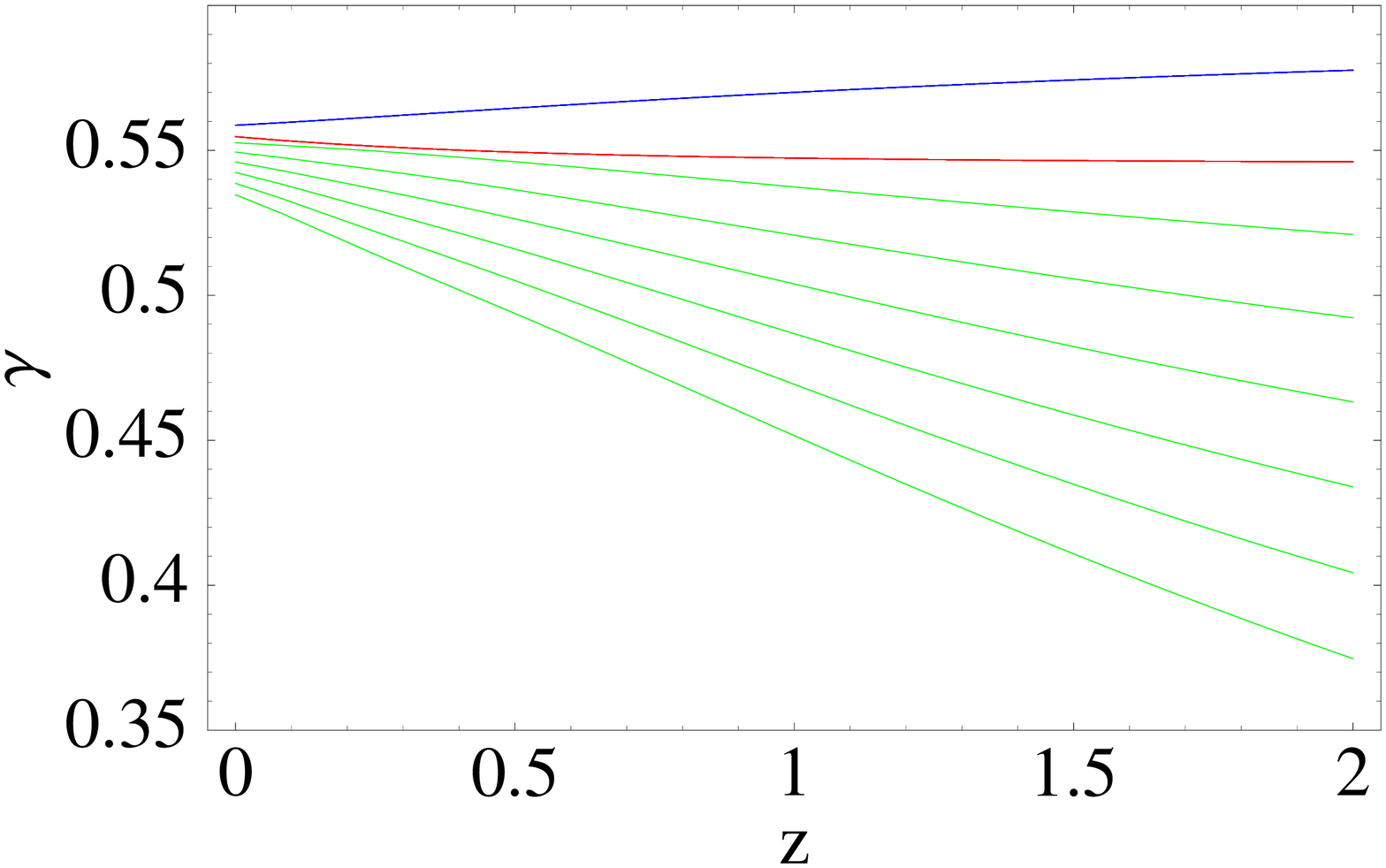}
\par\end{centering}

\caption{a) On the left, the function $f(z)$ is shown corresponding to the 
models of the right panel of figure \ref{R}. The growth factor is constrained by 
observations of the galaxy bias factor $b$ and the redshift distortion parameter 
$\bar\beta$ \cite{V02} via $f=b~\bar\beta$. The data are given here for reference. 
We see that one needs to go at redshifts $z>0.5$ in order to be able to distinguish 
between the different models. It is also clear that the present observations are 
not discrimative. Though not obvious from the behaviour of $f(z)$, the behaviour 
on small redshifts is very different as seen on the right panel.   
b) On the right, the function $\gamma(z)$ is displayed corresponding to the 
same models as on the left panel. But the order of the scalar-tensor models is 
reversed here, we have now from bottom to top $\frac{F_{\infty}}{F_0}=0.9301, 
0.94, 0.95, 0.96, 0.97, 0.98$. It is seen that $\gamma(z)$ has a quasi-linear 
behaviour on small redshifts and that large slopes can be obtained for 
scalar-tensor models when $C$ is close to $1$. Interestingly, this is also 
when the linear growth of perturbations is essentially the same as in 
$\Lambda$CDM for $z\gg 1$ in the matter dominated stage.} 
\lb{ST}
\end{figure}

\begin{figure}
\includegraphics[scale=.3]{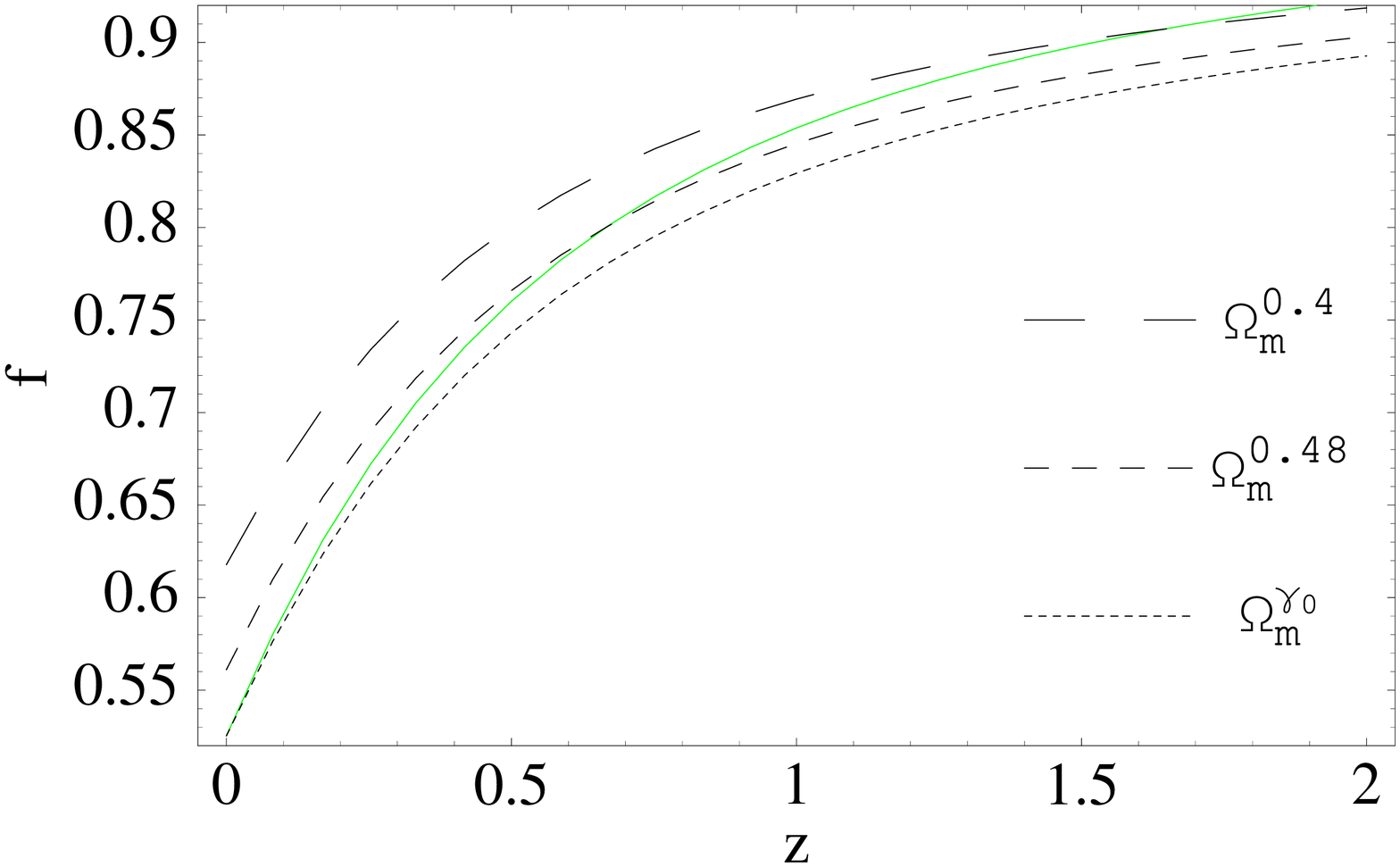}
\caption{
In green the growth function $f(z)$ is shown corresponding to the scalar-tensor 
model with $\frac{F_\infty}{F_0}=0.9301$ while the background is parametrised by 
$\Omega_{m,\infty}=0.93$ and  $\Omega_{m,0}=0.3$. This function is compared to 
three fits with a {\it constant} $\gamma:~0.4,~0.48,~\gamma_0=0.53$, where 
$\gamma_0$ is the value at $z=0$ of the true $\gamma(z)$. For $\gamma_0=0.53$, 
the fit is very good (by definition) on very small redshifts but bad on larger 
redshifts. Choosing $\gamma=0.4$ gives a good fit on large redshifts but is bad 
on small redshifts. Taking $\gamma=0.48$ yields a fit that is clearly different 
from $f$.} 
\end{figure}

All models shown in Figure \ref{ST} have exactly the same background expansion so 
that the difference in the growth of matter perturbations is solely due to the 
modification of gravity.
As we can see from Figure \ref{ST}, some models can be easily distinguished from each 
other using the growth of matter perturbations through the difference in the behaviour 
of $\gamma(z)$ on small redshifts. 
We note also for the models displayed in Figure \ref{ST} that the model inside GR 
is easily distinguished from the scalar-tensor models having a large slope $\gamma'_0$. 
For these ST models we have that $\gamma'_0$ is negative and it can be large, while it 
is (slightly) positive for the same background evolution inside GR. Actually as we can 
see from Figure 2 $\gamma'_0>0$ is true inside GR for the interesting range of 
cosmological parameters $\beta$ and $\Omega_{m,0}$, while we have generically 
$\gamma'_0<0$ for the interesting scalar-tensor models. Also, as already noted in 
\cite{GP07}, $|\gamma'_0|\lesssim 0.02$ inside GR while we see that it can be 
substantially larger outside GR. 
Obviously, when $|\gamma'_0|$ is large, assuming $\gamma=$ constant yields a poor 
approximation. 
Another interesting feature is the quasi-linear behaviour of $\gamma(z)$ 
on small redshifts $0\le z\le 0.5$. Such a behaviour could be probed observationnally and 
could allow to discriminate models whose parameter $\gamma_0$ are close to each other. 
The lower the value of $\gamma_0$, the better this potential resolution which improves 
as well when $w_{DE,0}$ decreases.  

This picture remains essentially the same for the best-fit models of Table 1. 
It is seen that the growth of matter perturbations on large redshifts is closer 
to the standard behaviour $\delta_m \propto a$ ($C=1$) in the best-fit 
scalar-tensor models of Table 1 than it is for the model inside 
GR with same parametrization of the background expansion. For the latter model, 
though a very pronounced phantom behaviour is needed today (see Table 1), it is 
not sufficient to make $\Omega_{m,\infty}$ higher than $0.95$ which has a 
crucial effect on the growth of matter perturbations on large redshifts.

To summarize, we have shown that our models have a characteristic signature in the growth 
of linear matter perturbations. On large redshifts inside the matter-dominated stage, we 
find a scaling behaviour for the matter perturbations which can substantially differ from 
$\Lambda$CDM and also from General Relativity (GR) with identical background evolution 
characterized, besides $\Omega_{m,0}$, by $\Omega_{m,\infty}$ or $\beta$. 
On small redshifts we find again a possible significant departure from $\Lambda$CDM 
\emph{and} models in General Relativity (GR) with same parametrization of the 
background expansion. 
Even for those models in which the growth of matter perturbations on large redshifts is 
close to that in $\Lambda$CDM, we find a large (negative) slope $\gamma'_0$, with $|\gamma'_0|$ 
much larger than in GR, whether $\Lambda$CDM or GR with an identical background evolution. 
For these models assuming a constant $\gamma$ would necessarily lead to a poor fit of the 
growth function $f$ (see Figure 4).
Such a behaviour on small redshifts would constitute a characteristic signature of our DE 
model being outside GR. Interestingly, those models that mimic $\Lambda$CDM on large redshifts 
are most easily distinguished from $\Lambda$CDM on small redshifts through their slope $\gamma'_0$. 
Though the results derived here are to some extent model dependent, it is clear that the growth 
of matter perturbations, especially when combined on small and large redshifts, can efficiently 
probe the nature of Dark Energy and in particular help in assessing whether we are dealing with 
a modified gravity DE model or not. 


\begin{thebibliography}{99}

\bibitem{P97} 
 S.J. Perlmutter et al., Ap. J. {\bf 483} 565 (1997), Nature {\bf 391} 51 (1998);
 A. G.~Riess, A. V.~Filippenko, P.~Challis {\it et al.}, Astron.~J. {\bf 116}, 1009 (1998);
 S. J.~Perlmutter, G.~Aldering, G.~Goldhaber {\it et al.}, Astroph.~J. {\bf 517}, 565 (1999);
 P.~Astier, J.~Guy, N.~Regnault {\it et al.}, Astron. Astroph. {\bf 447}, 31 (2006);

\bibitem{R07}
 Adam G. Riess {\it et al.}, Astrophys. J.{\bf 659}, 98 (2007), {\tt e-Print astro-ph/0611572};
 W.~M.~Wood-Vasey {\it et al.}  [ESSENCE Collaboration], Astrophys. J. {\bf 666}, 694 (2007).

\bibitem{SS00} 
 V. Sahni, A. A. Starobinsky, Int. J. Mod. Phys. D{\bf 9}, 373 (2000);   
 T. Padmanabhan, Phys. Rep. {\bf 380}, 235 (2003);
 E. J. Copeland, M. Sami and S. Tsujikawa, {\tt hep-th/0603057} (2006);
 V. Sahni, A. A. Starobinsky, Int. J. Mod. Phys. D{\bf 15}, 2105 (2006);
 R. Durrer, R. Maartens, Gen. Rel. Grav.{\bf 40}, 301 (2008);
 P. Ruiz-Lapuente, Class. Quant. Grav.{\bf 24}R91 (2007), {\tt e-Print 0704.1058}

\bibitem{Perc07} 
 W. J. Percival {\it et al.} {\tt e-Print 0705.3323}.

\bibitem{RPW88} 
 B. Ratra and P.J.E. Peebles, Phys. Rev. D{\bf 37} 3406 (1988);
 C. Wetterich, Nucl. Phys. B{\bf 302} 668 (1988).

\bibitem{D07}
 T.~M.~Davis {\it et al.},  Astrophys. J. {\bf 666}, 716 (2007).

\bibitem{ST} 
 N.~Bartolo and M.~Pietroni, Phys.\ Rev.\ D \textbf{61} 023518 (2000); 
 F.~Perrotta, C.~Baccigalupi and S.~Matarrese, Phys.\ Rev.\ D \textbf{61}, 023507 (2000); 
 Y. Fujii, K. Maeda, {\it The scalar-tensor theory of gravitation}, Cambridge Univ. Press (2003);
 V. Faraoni, {\it Cosmology in Scalar-Tensor Gravity}, Kluwer Academic, Dordrecht, 2004.

\bibitem{BEPS00} 
 B. Boisseau, G. Esposito-Far\`ese, D. Polarski and A.A. Starobinsky, 
 Phys. Rev. Lett. {\bf 85}, 2236 (2000).

\bibitem{EP01} 
 G. Esposito-Far\`{e}se and D. Polarski, Phys.\ Rev.\ D \textbf{63}, 063504 (2001).

\bibitem{GPRS06}
 R. Gannouji, D. Polarski, A. Ranquet, A. A. Starobinsky, {\it JCAP} {\bf 0609}, 016 (2006).

\bibitem{MSU06} 
 J. Martin, C. Schimd and J.-P.~Uzan, Phys.~Rev.~Lett.{\bf 96}, 061303 (2006).
 G. Barenboim, J. Lykken, {\tt e-Print 0711.3653};
 M. Demianski, E. Piedipalumbo, C. Rubano, P. Scudellaro, {\tt e-Print 0711.1043};
 S. Capozziello, P.K.S. Dunsby, E. Piedipalumbo, C. Rubano,  {\tt e-Print 0706.2615}; 


\bibitem{S98} 
 A. A. Starobinsky, JETP Lett. {\bf 68}, 757 (1998);

\bibitem{B06} 
 S. Bludman, {\tt e-Print astro-ph/0702085};
 D. Polarski, AIP Conf. Proc. 861, 1013 (2006), {\tt e-Print astro-ph/0605532};
 M. Ishak, A. Upadhye, D. N. Spergel, Phys. Rev. D {\bf 74}, 043513 (2006).

\bibitem{KST06}
 L. Knox, Y. S. Song, J. A. Tyson, Phys. Rev. D {\bf 74}, 023512 (2006);
 T. Chiba and R. Takahashi, Phys. Rev. D {\bf 75}, 101301 (2007);
 P. Zhang, M. Liguori, R. Bean, S. Dodelson, Phys.~Rev.~Lett.{\bf 99}, 141302 (2007). 

\bibitem{HKV07} 
 A. F. Heavens, T. D. Kitching, L. Verde, {\tt e-Print astro-ph/0703191}.

\bibitem{HL06} 
 D. Huterer, E. Linder, {\tt astro-ph/0608681}; 
 E. Linder, R. Cahn, {\tt astro-ph/0701317};
 S. Tsujikawa, {\tt e-Print 0705.1032};
 C. Di Porto, L. Amendola, {\tt e-Print 0707.2686}; 
 V. Acquaviva, L. Verde, {\tt e-Print 0709.0082}; 
 A. Kiakotou, O. Elgaroy, O. Lahav, {\tt e-Print 0709.0253};
 S.~Nesseris and L.~Perivolaropoulos, Phys. Rev. D {\bf 77}, 023504 (2008);
 Y.~Wang, {\tt e-Print 0710.3885};
 L. Hui, K. Parfree, {\tt e-Print 0712.1162}

\bibitem{FWZ05}
 Bo Feng, X.-L. Wang, X.-M. Zhang, Phys. Lett.B{\bf 607}, 35 (2005);
 Hao Wei, Rong-Gen Cai, Ding-Fang Zeng, Class. Quantum Grav.{\bf 22}, 3189 (2005).


\bibitem{GP07} R. Gannouji, D. Polarski, Phys. Lett. B 660, 439 (2008).

\bibitem{T02} D. Torres, Phys. Rev. D {\bf 66}, 043522 (2002).

\bibitem{LB07}
B. Li, John D. Barrow, Phys.Rev.D {\bf 75}, 084010 (2007). 

\bibitem{CP01} 
 M. Chevallier, D. Polarski, Int. J. Mod. Phys. D{\bf 10}, 213 (2001).

\bibitem{L03} 
 E. V. Linder, Phys. Rev. Lett. {\bf 90}, 091301 (2003).

\bibitem{PC03} 
 E.~Di Pietro and J.~F.~Claeskens, Mon.\ Not.\ Roy.\ Astron.\ Soc.\  {\bf 341}, 1299 (2003);
 S.~Nesseris and L.~Perivolaropoulos, Phys.\ Rev.\  D {\bf 70}, 043531 (2004); 
 R.~Lazkoz, S.~Nesseris and L.~Perivolaropoulos, JCAP {\bf 0511}, 010 (2005).

\bibitem{RU02}  
 E.~Garcia-Berro, E.~Gaztanaga, J.~Isern, O.~Benvenuto and L.~Althaus, {\tt e-Print astro-ph/9907440};
 L. Amendola, P. S. Corasaniti, F. Occhionero, {\tt astro-ph/9907222}; 
 A.~Riazuelo and J.~P.~Uzan, Phys.\ Rev.\  D {\bf 66}, 023525 (2002);

\bibitem{E05}
 D.~J.~Eisenstein {\it et al.}  [SDSS Collaboration], Astrophys.\ J.\  {\bf 633}, 560 (2005).

\bibitem{WM06}
 Y.~Wang and P.~Mukherjee, Astrophys.\ J.\  {\bf 650}, 1 (2006).

\bibitem{N07}
I. P. Neupane, e-Print: {\tt e-Print 0711.3234}.

\bibitem{KS07} 
 M. Kunz, D. Sapone, Phys. Rev. Lett. {\bf 98}, 121301 (2007).

\bibitem{P84} 
 P. J. E. Peebles, Astrophys. J.{\bf 284}, 439 (1984).
\bibitem{LLPR91} 
 O. Lahav, P. B. Lilje, J. R. Primack, M. J. Rees, MNRAS{\bf 251}, 128 (1991).

\bibitem{WS98} 
 L. Wang, P. J. Steinhardt, Astrophys. J.{\bf 508}, 483 (1998).

\bibitem{V02}
L.~Verde {\it et al.}, Mon.\ Not.\ Roy.\ Astron.\ Soc.\  {\bf 335}, 432 (2002); 
E.~Hawkins {\it et al.}, Mon.\ Not.\ Roy.\ Astron.\ Soc.\  {\bf 346}, 78 (2003); 
M.~Tegmark {\it et al.}  [SDSS Collaboration], Phys.\ Rev.\  D {\bf 74}, 123507 (2006); 
N.~P.~Ross {\it et al.}, arXiv:astro-ph/0612400; 
J.~da Angela {\it et al.}, arXiv:astro-ph/0612401.

\end{thebibliography}
\end{document}